\documentclass{PoS} 
\usepackage{sidecap} 
\usepackage{wrapfig}

\def \HI {H{\sc \,i}}
\def \WpHz {W Hz$^{-1}$}
\def\MOLH {\hbox{${\rm H}_2$}}  
\def\scm  {$\hbox{{\rm cm}}^{-2}$}
\def \AL {$\alpha $}
\def\lapp{\ifmmode\stackrel{<}{_{\sim}}\else$\stackrel{<}{_{\sim}}$\fi}
\def\gapp{\ifmmode\stackrel{>}{_{\sim}}\else$\stackrel{>}{_{\sim}}$\fi}

\title{Blind Wide Area Surveys: Where will we find redshifted atomic and molecular absorption?}

\ShortTitle{Blind Wide Area Surveys: Redshifted atomic and molecular absorption}

\author{\speaker{Stephen J. Curran}%
  \\
        University of New South Wales\\
        E-mail: \email{sjc@phys.unsw.edu.au}}

\author{Matthew T. Whiting\\
        CSIRO Australia Telescope National Facility\\
        E-mail: \email{Matthew.Whiting@csiro.au}}

\author{John K. Webb\\
  University of New South Wales\\
        E-mail: \email{jkw@phys.unsw.edu.au}}

\abstract{Spectroscopy of redshifted radio absorption of atomic and
  molecular species provide excellent probes of the cold component of the 
  gas in the early Universe which can be used to address many
  important issues, such as measuring baryonic content, probing
  large-scale structure and galaxy evolution, as well as obtaining
  independent measurements of various combinations of fundamental
  constants at large look-back times. However, such systems are
  currently very rare with only 80 detected in \HI\ 21-cm and five in
  OH and millimetre-band species. Here we summarise the main selection
  criteria responsible for this and how the next generation of radio
  telescopes are expected to circumvent these through their 
  wide instantaneous bandwidths and fields-of-view. Specifically:
  \begin{enumerate}
    \item {\em \HI\ in absorbers occulting distant quasars:}
      These are usually found in known optical absorbers and wideband
      radio surveys could reveal a much fainter population. However,
      the 21-cm absorption strength may be correlated with the width
      of the singly ionised metal species, suggesting that these may
      be weak, and due to the effects of a flat expanding Universe on
      the covering factor, we expect the highest detection rates at $z
      < 1$.
      
    \item {\em \HI\ absorption associated with the host
      galaxy/quasar:} Due to high degrees of ionisation/excitation
      rendering 21-cm undetectable near active nuclei with
      ultra-violet luminosities of $L_{\rm UV}\gapp10^{23}$ \WpHz,
      future searches should be magnitude limited, e.g. at $z>1$, blue
      magnitudes should be $B \gapp19$ with $z>2 \Rightarrow B
      \gapp21$ and $z>3 \Rightarrow B \gapp22$.

       \item {\em OH (and millimetre-band) absorption:} For all of the
         known redshifted molecular absorption systems a correlation
         is found between the molecular fraction and the
         optical--near-infrared colour ($V-K$), with the five known OH
         absorbers all having $V-K\gapp5$. Therefore spectral scans towards
         extremely red radio sources are expected to uncover any dusty intervening,
         molecular rich absorbers reponsible for the obscuration of the optical light.
       
    \end{enumerate}
}
\FullConference{Panoramic Radio Astronomy: Wide-field 1-2 GHz research on galaxy evolution\\
                 June 2-5 2009\\
                 Groningen, the Netherlands}

\begin{document}

\section{Optically Selected Intervening Absorbers}

With total neutral hydrogen column densities of $N_{\rm
  HI}\gapp10^{20}$ \scm\ and precisely determined redshifts, the
detection of \HI\ 21-cm and OH in damped Lyman-$\alpha$ absorption
systems (DLAs) should be like shooting fish in a barrel. However,
there are only 40 cases, out of 151 published searches (see
\cite{cur09a}), of 21-cm being detected in redshifted intervening
absorbers (Lyman-$\alpha$ and Mg{\sc \,ii}) and there has never been a
detection of molecular absorption in either the decimetre (OH) or
millimetre band (CO, HCO$^+$, etc., summarised in \cite{cmpw03}).

\subsection{Absorption by Atomic Gas}

In an absorption system the neutral hydrogen column density is related
to the 21-cm line strength via
$N_{\rm HI}=1.823\times10^{18}\,T_{\rm spin}\int\!\tau\,dv\,,$
where $T_{\rm spin}$  is the mean harmonic
spin temperature of the gas and $\int\!\tau\,dv$ is the velocity
integrated optical depth of the line. The optical depth is defined via
$\tau\equiv-\ln\left(1-\frac{\sigma}{f\,S}\right)$, where $\sigma$ is
the depth of the line and $S$ and $f$ the flux density and covering
factor of the background continuum source, respectively. Therefore
in the optically thin regime, the  expression simplifies to
$N_{\rm HI}=1.823\times10^{18}\frac{T_{\rm spin}}{f}\int\!\frac{\sigma}{S}\,dv$
and if the Lyman-$\alpha$ (from which $N_{\rm HI}$ is determined) and
21-cm absorption arise in the same sight-line, the velocity integrated
``optical depth'', $\int\!\frac {\sigma}{S}\,dv$, gives the ratio of
the spin temperature to the covering factor, $T_{\rm spin}/f$.

Since the 21-cm detections occur overwhelmingly at $z_{\rm
  abs}\lapp1$, where there is a mix of detections and non-detections,
\cite{kc02} advocate a scenario where low redshift absorbers have a
mix of spin temperatures, while those at high redshift exhibit
exclusively high temperatures. This result, however, relies on the
$z_{\rm abs}>2$ absorbers having the maximum permissible covering
factor ($f=1$), which yields the maximum possible spin temperature
\cite{cmp+03}. This results in a huge disparity in the values,
specifically $T_{\rm spin}\approx200 {\rm ~to}\,\gapp9\,000$~K
\cite{kc02}. Furthermore, with a maximum possible spin temperature of
$T_{\rm spin}/(f=1)\approx140$~K at $z_{\rm abs}=2.289$ \cite{ykep07},
the DLA towards TXS 0311+430 rules out the hypothesis that a high
redshift necessarily entails a high spin temperature.

\begin{figure}[h]
\centering
\includegraphics[angle=270,scale=0.62]{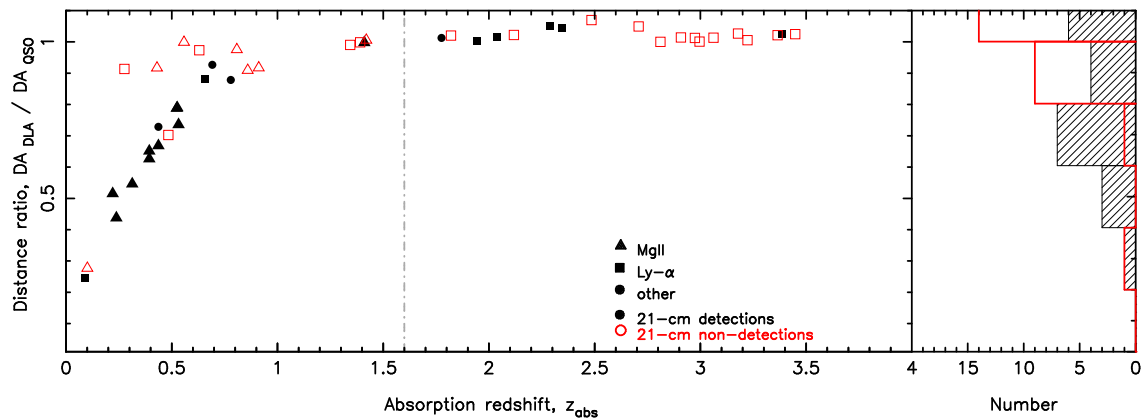}
\caption{The angular diameter distance ratio versus redshift
for the DLAs searched in 21-cm absorption. The filled symbols/hatched histogram
 represent the 21-cm detections and the
  unfilled symbols/unfilled histogram the non-detections. 
The symbol shapes designate the transition through which 
the absorber was discovered. Updated from \cite{cw06}.}
\label{distance-z}
\end{figure}
Although without high resolution imaging of the absorber {\em and} the
background emission region at the redshifted 21-cm frequency the
$T_{\rm spin}/f$ degeneracy cannot be broken, we can 
suggest that the covering factor may nonetheless play an important r\^{o}le.
This is apparent in Fig. \ref{distance-z}, where we show the ratio of the
angular diameter distances to the absorber and quasar against redshift:
Due to the effects of a flat expanding Universe, beyond $z\approx1.6$ the
angular diameter distance decreases with redshift and so, whereas low redshift
systems may exhibit a variety of $DA_{\rm DLA}/DA_{\rm QSO}$ ratios, high redshift
absorbers will {\em always} have $DA_{\rm DLA}/DA_{\rm QSO}\approx1$.

That is, at $z_{\rm abs}\gapp1$ the absorber and quasar are
essentially at the same angular diameter distance and so for given
absorber and emitter cross-sections, these absorbers will cover the
background emission much less effectively that those with low $DA_{\rm
  DLA}/DA_{\rm QSO}$ ratios. The mix at low redshift arises from the
mix of low and high redshift background quasars, with the overall
distribution following the same pattern as that of the ``spin temperature'' distribution
of \cite{kc02} -- a variety at low redshift and exclusively high values at
high redshift. Although in this case it is a covering factor effect,
which is not subject to the same assumptions.

This can also explain the high detection rate in DLAs originally found
through the Mg{\sc \,ii}, rather than the Lyman-\AL, line: The Mg{\sc
  \,ii} doublet can be observed by ground-based telescopes at redshifts
of $0.2\lapp z_{\rm abs}\lapp2.2$, although the Lyman-\AL\ transition is restricted
to $z_{\rm abs}\gapp1.7$. That is, Mg{\sc \,ii} observations preferentially
select low redshift absorbers, thus not resulting in exclusively high 
$DA_{\rm DLA}/DA_{\rm QSO}$ ratios (and hence lower covering factors).
\begin{figure}[h]
\centering
\includegraphics[angle=270,scale=0.68]{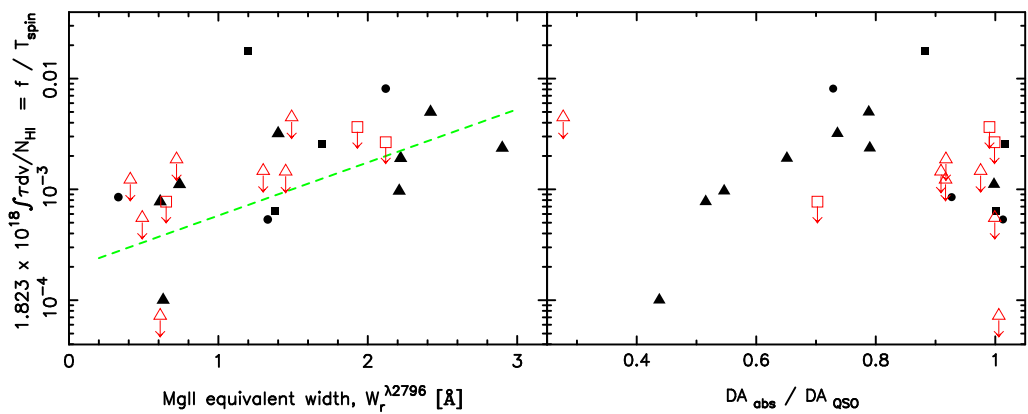}
\caption{The 21-cm line strength versus the rest frame
  equivalent width of the Mg{\sc \,ii} 2796 \AA\ line (left) and the
  angular diameter distance ratio (right) for the DLAs searched in
  21-absorption. The correlation shown by the least-squares fit
in the left panel is significant at $2.52\sigma$ (see \cite{cur09a}).}
\label{strength_W_ratio}
\end{figure}
On the subject of Mg{\sc \,ii} absorption, \cite{ctp+07} find a
correlation between the 21-cm line strength and the Mg{\sc \,ii} 2796
\AA\ equivalent width, although there are also DLAs which span similar
equivalent widths while remaining undetected in 21-cm
(Fig. \ref{strength_W_ratio}, left). However, of the nine which have
good limits, eight of the non-detections have $DA_{\rm DLA}/DA_{\rm
  QSO}\gapp0.9$ (Fig. \ref{strength_W_ratio}, right), again suggesting
a strong covering factor effect due to the line-of-sight geometry.

\subsection{Absorption by Molecular Gas}
\label{abmg}

\begin{figure}[h]
\centering \includegraphics[angle=0,scale=0.62]{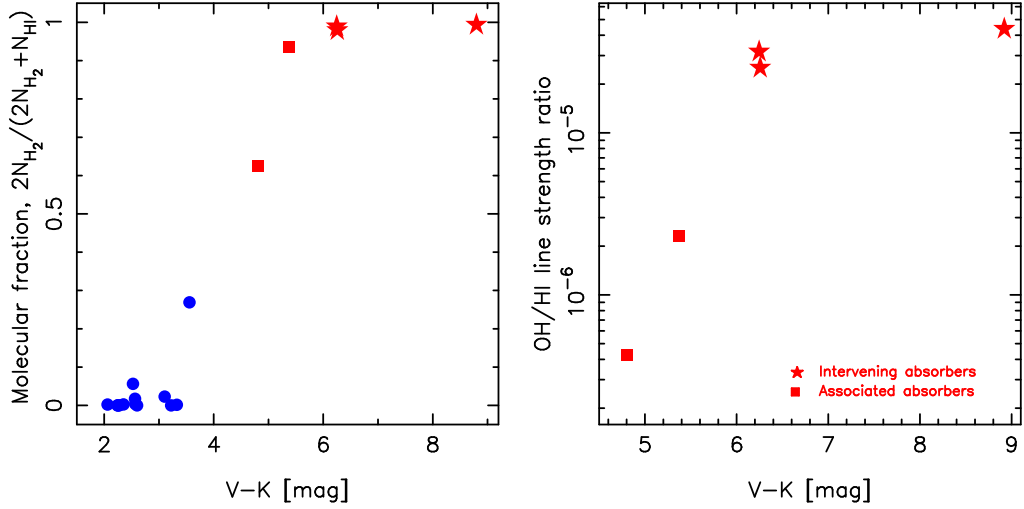}
\caption{The molecular fraction [${\cal F}$] (left) and the normalised OH line
  strength [$2.38\times10^{14}\int \tau_{_{\rm OH}}\,
    dv/1.82\times10^{18}\int\tau_{_{\rm HI}}\, dv$] (right) versus the
  optical--near infrared colour. The circles represent the \MOLH-bearing
DLAs (all intervening absorbers) and the squares and stars the OH absorbers.
Updated from \cite{cwm+06}, where the correlation between ${\cal F}$ and $V-K$
is now significant at $3.7\sigma$.}
\label{frac-col}
\end{figure}

Although the \MOLH\ molecule has been detected through the Lyman and
Werner ultra-violet bands\footnote{Redshifted into the optical-band at
  $z_{\rm abs}\gapp1.7$.} in 17 DLAs, extensive millimetre-wave band
searches have yet to yield a detection, despite exceeding the
sensitivities to required to detect the known millimetre/OH absorbers
by an order of magnitude (\cite{cmpw03} and references therein).  We
have noted \cite{cwm+06}, however, that these radio-band absorbers
have molecular fractions ${\cal F}\equiv\frac{2N_{\rm H_2}}{2N_{\rm H_2}+N_{\rm
    HI}}\approx0.7 - 1$ and optical--near-infrared colours of
$V-K\gapp5$, whereas the optical-band absorbers have ${\cal F}\sim10^{-7} - 0.3$ and
$V-K\lapp4$ (Fig.~\ref{frac-col}).
This is strong evidence that quasar light is reddened by the
dust in the foreground absorber, which is necessary to prevent the
dissociation of the molecular gas by the ambient ultra-violet field. 

\section{Radio Selected Absorbers}
\subsection{Intervening Absorption}

From the above discussion (Sect. \ref{abmg}), it is apparent that DLAs
are simply too blue to indicate a sufficiently large column of dust
conducive to molecular abundances which can be detected by current
radio telescopes, although the SKA will be capable of detecting
normalised OH column densities of $\frac{N_{\rm OH}}{N_{\rm HI}}\lapp
10^{-6.5}\,\frac{f_{\rm OH}}{f_{\rm HI}}\,\frac{T_{\rm spin}}{T_{\rm
    ex}}$ (at $V-K\lapp5$, Fig. ~\ref{frac-col}). In the meantime,
searches for 18-cm OH absorption should be targetted towards very
red, radio-loud objects. However, this very redness means that
there are generally no optical redshifts of any absorption
features towards these quasars, which in turn means that
we have no knowledge of which frequency to tune the receivers.

Therefore spectral scans of the entire redshift space towards red
quasars are required in order to detect the absorber responsible for
the obscuration of the optical light. Although they will not be any
more sensitive than current state-of-the-art radio telescopes, the
large instantaneous bandwidths (e.g. 300 MHz with the ASKAP) will make
the forthcoming SKA pathfinders ideal instruments with which to
undertake such spectral scans. Furthermore, the large fields-of-view
(e.g. 30 square degrees with the ASKAP) will facilitate truly blind
surveys, where full range spectra ($0 \leq z_{_{\rm HI}} \lapp1.0$ and
$0\leq z_{_{\rm OH}} \lapp1.4$) of large areas of sky may be obtained in a
few re-tunings.

\subsection{Associated Absorption}

Although we may have to wait for the next generation of radio
telescopes to undertake effective full spectral scans, we can currently use the
redshift of the background radio source to search for absorption
associated with the host galaxy/quasar.
\begin{figure}[h]
\centering
\includegraphics[angle=270,scale=0.62]{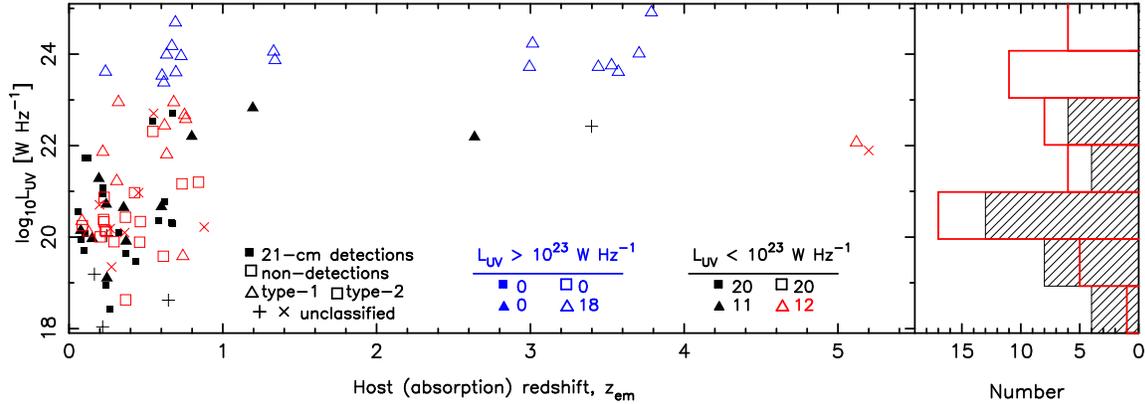}
\caption{The $\lambda=1216$ \AA\ luminosity versus the redshift for
  the radio galaxies and quasars searched in associated 21-cm
  absorption. The symbol and histogram filling is as per
  Fig. \protect\ref{distance-z} with the shapes representing the AGN
  classifications:  Triangles -- type-1, squares --
  type-2 and {\sf \large +} \& {\sf x} designating undetermined types for
  the detections and non-detections, respectively. Updated from
  \protect\cite{cww+08}.}
\label{lum-z}
\end{figure}
However, from a recent survey of $z_{\rm em}\approx3 - 4$ radio
sources for \HI\ and OH, we did not detect any absorption in the hosts
of any of these objects. Upon a thorough analysis of the optical
photometry, we found that all of our targets have ultra-violet
luminosities of $L_{\rm UV}\gapp10^{23}$ \WpHz\ \cite{cww+08}. This
could be interpreted as the by-product of some other selection effect
which causes the paucity of \HI\ 21-cm absorption\footnote{In light of
  the \HI\ non-detections it is not surprising that we did not detect
  OH. In any case, all of the targets have $V-K\leq2.6$, suggesting a
  low level of dust obscuration and thus low molecular fractions
  (Sect. \ref{abmg}).} at high redshift, if it were not for the fact
we found this to apply at {\em all} redshifts (Fig. \ref{lum-z}).

It is therefore apparent that we have identified a critical ultra-violet
luminosity, above which the gas is excited to beyond the detection
limits of current instruments. Higher UV luminosities can also account
for the lower 21-cm detection rates in quasars in comparison to radio
galaxies \cite{cww08l}, which is currently attributed to unified
schemes of AGN, where quasars are predominantly type-1 objects and
radio galaxies type-2 (e.g. \cite{gs06a} and references therein). Although,
all of the $L_{\rm UV}\gapp10^{23}$ \WpHz\ are indeed type-1, it is
the inclusion of these which gives the apparent bias against type-1
objects, with a 50\% 21-cm detection rate being found for {\em both}
AGN types at $L_{\rm UV}\lapp10^{23}$ \WpHz\ (Fig. \ref{lum-z}).  Note
also that, with the exclusion of the $L_{\rm UV}\gapp10^{23}$
\WpHz\ sources, the detection rate in ``compact objects'' (gigahertz
peaked spectrum and compact steep spectrum sources) is not any higher
than for the remainder of the radio/galaxy sample \cite{cww08l}.

\begin{figure}[h]
\centering
\includegraphics[angle=270,scale=0.62]{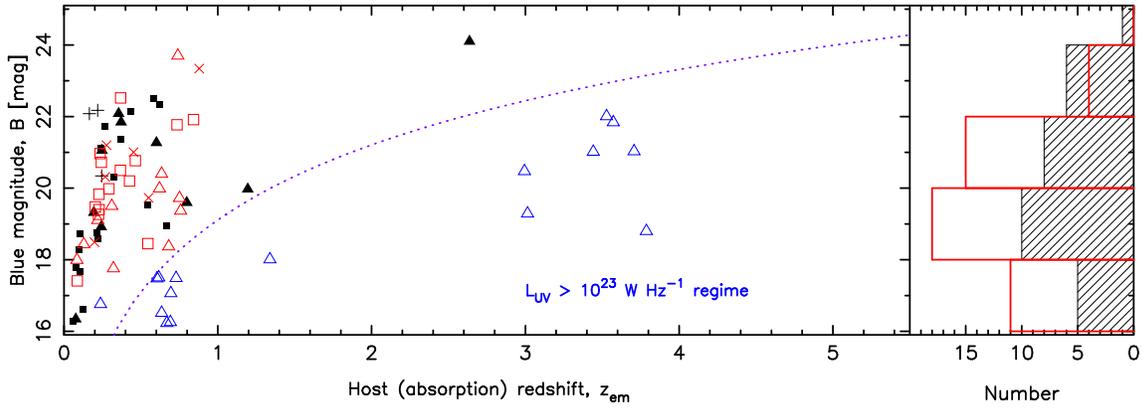}
\caption{As Fig. \protect\ref{lum-z}, but showing the blue magnitude of the source on the ordinate.
The curve shows which $B$ magnitude corresponds to $L_{\rm UV}=10^{23}$ \WpHz\ for a spectral slope
of $\alpha = -1.5$.}
\label{B-z}
\end{figure}

Although the detection rate is only 50\% at $L_{\rm UV}\lapp10^{23}$
\WpHz\ (probably due to the orientation of the galactic disk,
\cite{cww08l}), we believe  that searches for associated absorption
will have to be restricted to these luminosities if detection rates of
$>0$\% are to be obtained. As a search diagnostic, in Fig. \ref{B-z}
we show the blue magnitude against the redshift for all of the
published $z_{\rm em}\gapp0.1$ searches.
From this we see that, although our previous $z_{\rm em}\approx3 - 4$
survey was generally restricted to $B\gapp19$, these magnitudes are
still not sufficiently faint to indicate $L_{\rm UV}\lapp10^{23}$
\WpHz\ at such high luminosity distances. That is, once again,
optical selection of targets introduces a bias against the detection
of radio absorption lines, and blind wide-band searches over large
fields-of-view are expected to significantly increase the number of
\HI\ and OH absorbers at high redshift.

\end{document}